\documentclass[aps,prb,onecolumn,nofootinbib,citeautoscript,10pt]{revtex4-2}  
\usepackage{amsmath,amssymb,bm} 
\usepackage{graphicx}

\usepackage[tight]{subfigure} 

\usepackage{color} 
\usepackage[papersize={8.5in,11in}]{geometry}
\usepackage[colorlinks=true]{hyperref}
\hypersetup{
    bookmarks=true,         
    unicode=false,          
    pdftoolbar=true,        
    pdfmenubar=true,        
    pdffitwindow=false,     
    pdfstartview={FitH},    
    pdfkeywords={keyword1} {key2} {key3}, 
    pdfnewwindow=true,      
    colorlinks=true,       
    linkcolor=magenta, 
    citecolor=blue,        
    filecolor=magenta,      
    urlcolor=blue           
} 

\usepackage{amsfonts,bbold}
\usepackage{upgreek}
\usepackage{slashed}
\usepackage{latexsym}
\usepackage{ulem}

\usepackage{wrapfig} 
\usepackage{wasysym}			
\usepackage{amsthm}
\usepackage{xcolor}
\usepackage{hyperref} 
\usepackage{breakurl}

\def \nn{\nonumber \\}

\begin{document}

\title{Non-Hermitian generalizations of the Yao-Lee model augmented by SO(3)-symmetry-breaking terms}

\author{Ipsita Mandal}

\affiliation{Department of Physics, Shiv Nadar Institution of Eminence (SNIoE), Gautam Buddha Nagar, Uttar Pradesh 201314, India
\\and\\
Freiburg Institute for Advanced Studies (FRIAS), University of Freiburg, D-79104 Freiburg, Germany}

\begin{abstract}
We investigate non-Hermitian versions of the Yao-Lee model, supplemented by various kinds of SO(3)-symmetry-breaking terms, preserving the solvability of the model. The parent model hosts three species of Majorana fermions, thereby serving as an extension of the two-dimensional Kitaev model on the honeycomb lattice. The non-Hermitian couplings represent generic situations when the system is coupled to the environment and, thus, undergoes dissipation. The resulting eigenvalue spectrum and the eigenmodes show a rich structure of exceptional points as well as non-Hermitian skin effects. We chart out such exotic behaviour for some representative parameter regimes.
\end{abstract}

\maketitle

\tableofcontents

\section{Introduction}

The two-dimensional (2d) Kitaev model on a honeycomb lattice \cite{kitaev} is one of the rare examples of a solvable system harbouring quantum spin liquids. A spin-orbital generalization of the Kitaev model is the SU(2)-symmetric spin liquid model, proposed by Yao and Lee \cite{yao-lee}, where the effective degrees of freedom are three flavours of Majorana fermion species having degenerate energy eigenvalues. In this paper, we couple this Yao-Lee model, supplemented by terms which break the degeneracy of its spectrum, to the environment, which then represents a dissipative system \cite{rev-emil}. Motivated by the construction used in Ref.~\cite{kang-emil}, we employ a non-Hermitian description for the effective description of the resulting behaviour. The aim is to uncover some unusual features related to exceptional points and the non-Hermitian skin effect (NHSE). The latter refers to the emergence of localized eigenstates \cite{hatano-nelson, hatano-nelson2, lee, Yao2018, slager, Okuma2020, Kunst2018,Martinez2018,elisabet,Xiao2020,okuma2023,Lin2023, maria_emil}, with no familiar counterpart in generic Hermitian systems.

Exceptional points (EPs) are singular points of generic complex matrices at which multiple eigenvalues, along with their eigenvectors, coalesce \cite{Berry2004,Heiss_2012,PhysRevX.6.021007,ep-optics,ozdemir2019parity}. Their appearance is intimately connected to topological phases \cite{ips-ep-epl,ips-tewari,rev-emil,kang-emil,ips-prl,ips-kang}. The investigations of the non-Hermitian Kitaev spin liquid model in Refs.~\cite{kang-emil,yang2022ep} outlined the effects of non-Hermiticity, whereby the Dirac points of the emergent Majorana fermions of the original (Hermitian) model split into pairs of exceptional points, with Fermi arcs connecting them. The authors also discussed the NHSE, characterized by the accumulation of the eigenstates of the lattice of a ``non-Hermitian Hamiltonian'' onto the lattice-boundaries \cite{hatano-nelson,hatano-nelson2, lee, Yao2018, Okuma2020, Kunst2018,Martinez2018,elisabet, Xiao2020,okuma2023,Lin2023,maria_emil}. In this paper, we extend the explorations in Ref.~\cite{kang-emil} by considering a non-Hermitian version of the Yao-Lee model (for example, Ref.~\cite{ips-kang}), which is incorporated via a decorated honeycomb lattice, and which can support three copies of the Majorana fermion species found in the 2d Kitaev model. In order to obtain distinct nontrivial phenomenology, we break the degeneracy among the three Majorana species by adding inter-species interactions like diagonal $K$-terms \cite{sreejith}, off-diagonal $\Gamma$-terms \cite{sreejith}, and Dzyaloshinskii-Moriya interaction (DMI) terms in presence of an external magnetic field $ \pmb{\mathcal B} $ \cite{Banerjee_NatPhys2013,sreejith,akram}. Such a system is expected to harbour a richer phase diagram than the cases studied in Ref.~\cite{kang-emil}.

An intriguing aspect of non-Hermitian systems is the fact that the periodic boundary condition (PBC) and the open boundary condition (OBC) spectra are completely different, with no familiar notions of bulk-boundary correspondence (which exist for Hermitian systems). This is in stark contrast with the fact that boundary conditions do not affect the bulk spectra for Hermitian systems in the thermodynamic limit, except for the additional edge states appearing for OBCs. The inherent difference in the PBC and OBC spectra is the root cause for the emergence of the NHSE. The anomalous localization phenomena can be resolved using the biorthogonal bulk-boundary correspondence, which also helps us formulate real-space topological invariants capturing the number of edge states for OBCs \cite{Kunst2018, ips-bp}. Although the focus of this paper is not to determine such invariants, we will highlight how non-Hermitian systems display a strong sensitivity to boundary conditions by considering zigzag boundaries in decorated honeycomb lattices.

In the context of the NHSE, while most of the existing literature focusses on one-dimensional (1d) lattices \cite{Kunst2018,elisabet,paolo,maria_emil}, here we consider two-dimensional (2d) lattices harbouring six bands. We will demonstrate some examples of the variety of skin effects. and localization/delocalization of various eigenstates, that can arise in non-Hermitian versions of the Yao-Lee model (for example, Ref.~\cite{ips-kang}), whose threefold degenerate spectrum is broken by additional terms. We will see that the different kinds of symmetry-breaking terms affect the emergence conditions for the NHSE differently, which is tied to the condition whether those terms lead to a mixing of the three Majorana species. For example, the condition for the appearance of the NHSE for the $K$-terms turns out to follow the same arguments as found in Ref.~\cite{kang-emil}, because no inter-species interaction is caused by the $K$-couplings. On the contrary, the remaining two scenarios explored in this paper cause nontrovial interactions between the three species, opening up large parameter ranges for the NHSE to appear, which are beyond those found in Ref.~\cite{kang-emil}.

 The paper is organized as follows. In Sec.~\ref{secpbc}, we review the construction of the original Yao-Lee model featuring three emergent Majorana fermion species. In Sec.~\ref{secdiagk} and Sec.~\ref{secgam}, we add nearest-neighbour exchange interactions, which are flavour-diagonal and flavour-off-diagonal, respectively. Sec.~\ref{secmag} deals with the effects of magnetic fields and DMI on the Yao-Lee Hamiltonian. Finally, we end with a summary and outlook in Sec.~\ref{secsum}.

\begin{figure}[]
	\centering
\includegraphics[width=0.25 \textwidth]{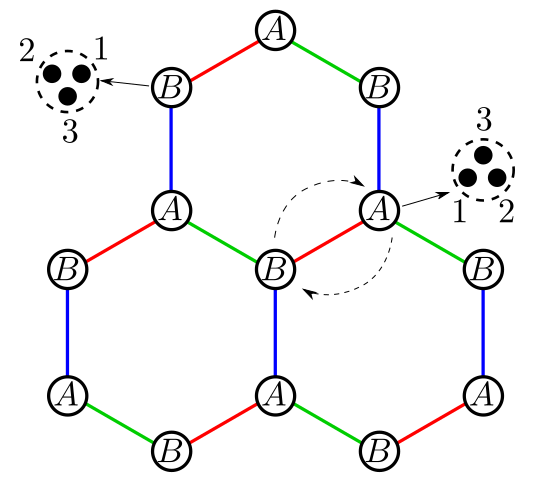}
\caption{\label{fighoney}
The schematic representation of the decorated honeycomb lattice \cite{yao-lee}. The sublattice sites of the honeycomb lattice are labelled by $A$ and $B$, each of which hosts an equilateral triangle. The three sites at the vertices each triangle are indicated by $\lbrace 1,2,3 \rbrace $. The types of the inter-triangle bonds are denoted by $x$-link (green), $y$-link (red), and $z$-link (blue).
}
\end{figure}

\section{Decorated honeycomb lattice: Yao-Lee model}
\label{secpbc}

We consider an extension of the original Kitaev model \cite{kitaev} on the honeycomb lattice, known as the Yao-Lee model \cite{yao-lee}, constructed as a \textit{decorated honeycomb lattice}, as shown schematically in Fig.~\ref{fighoney}. It is a 2d system consisting of spin-$1/2$ moments, localized at the vertices of equilateral triangles, whose centres are located at the sublattice sites of a honeycomb lattice.
This decorated honeycomb lattice is also known as the star or the 3--12 lattice.  The corresponding Hamiltonian 
is given by
\begin{align}
\label{hamyl}
    H_{YL} = \sum_{ \langle j l \rangle_{\alpha {\rm -link}}} 
    J_\alpha \left[ \tau_{j}^{(\alpha)} \, \tau_{l}^{(\alpha)}\right ]
\left( \pmb{\sigma}_{j} \cdot \pmb{\sigma}_{l} \right), \quad \alpha \in \{x,y,z\} \,,
\end{align}
where $ J_\alpha$ represents the nearest-neighbor hopping strength for the $\alpha$-type link (with $\mathbf J \equiv \lbrace J_x, \, J_y, \, J_z \rbrace $). The vector operators $ \boldsymbol \tau_{j} 
\equiv \lbrace \tau_{j}^{(x)} , \tau_{j}^{(y)}, \tau_{j}^{(z) }\rbrace$ and $ \boldsymbol \sigma_{j} 
\equiv \lbrace \sigma_{j}^{(x)} , \sigma_{j}^{(y)}, \sigma_{j}^{(z) }\rbrace$, defined on the site $ j $, consist of the three spin-$1/2$ operators (Pauli matrices) as their components. They act on the orbital and spin degrees of freedom on the $j^{\rm th}$-site, respectively.

The plaquette flux operator, defined as $W= \tau_i^x \, \tau_j^y \, \tau_k^z\, \tau_l^x\, \tau_m^y\,\tau_n^z\otimes \mathbb{1}$, has eigenvalues $\pm 1$. Since it commutes with $H_{YL} $, the eigenstates of $H_{YL}$ can be labelled by the eigenvalues of $W$.
Analogous to the treatment of the original Kitaev's model, the Hamiltonian is rewritten by introducing Majorana fermion operators for the spin operators in an extended Fock space as follows:
\begin{align}
\sigma_{j}^{(\alpha)} = -\frac{ i } {2} \,\epsilon^{\alpha \beta \gamma}c_{j}^{(\beta)}c_{j}^{(\gamma)} \,,
\quad \tau_{j}^{(\alpha)} = - \frac{ i } {2}  \,\epsilon^{\alpha \beta \gamma} \,
d_{j}^{(\beta)} \,d_{j}^{(\gamma)} \,, \quad
\sigma_j^\alpha \,\tau_j^\beta = i \,c_{j}^{(\alpha)} \, d_j^{(\beta)} \,.
\end{align}
This fermionisation procedure allows us to solve the system exactly, with the physical states obtained via the projection operator $P=\prod_j (1+D_{j})/2$, where
$ D_{j} = -i \,c_{j}^{(x)}\, c_{j}^{(y)} \, c_{j}^{(z)} \,d_{j}^{(x)} \,d_{j}^{(y)} \,d_{j}^{(z)}$.
The original Hamiltonian is related to the Hamiltonian $\mathcal H_{YL} $ in the Majorana representation via $ H_{YL}=  P \,\mathcal{H}_{YL}\, P$, where
\begin{align}
    \mathcal{H}_{YL} & =
 i  \sum_{\langle jl \rangle_{\alpha-{\rm links}} }
  J_\alpha \, u^{\alpha}_{jl}
    \left [ c_{j}^{(x)} \, c_{l}^{(x)} +  c_{j}^{(y)} \,c_{l}^{(y)} 
    +   c_{j}^{(z)} \, c_{l}^{(z)} \right ] ,
\end{align}
and $u^{\alpha}_{ jl}=-i\, d_{j}^{(\alpha)}\, d_{l}^{(\alpha)}$ are the bond operators with eigenvalues $\pm 1 $. There are three Majorana fermion species, labelled by $x$, $y$, and $z$, which reside at the vertices $1$, $2$, and $3$, respectively, of each equilateral triangle. 
Because the $u^{\alpha}_{ jl}$'s commute with the Hamiltonian as well as among themselves, all the eigenstates of $ \mathcal{H}_{YL} $ can be labelled by the eigenvalues of $u^{\alpha}_{ jl} $. In particular, the ground state corresponds to setting $u^{\alpha}_{ jl} = 1 $ (or, equivalently, $u^{\alpha}_{ jl} = -1$), which is in fact a vortex-free state (or zero-flux sector).

For $ \mathcal{H}_{YL} $ with Hermitian couplings, the spectrum of each species of Majorana fermions is the same as that in the Kitaev model and, effectively, we get three decoupled copies of the Kitaev model spectrum as a consequence of the global SO(3) symmetry. Let us focus on the zero-flux sector, where we can choose $u^{\alpha}_{ jl} = 1 $. Implementing this choice, the periodic Hamiltonian can be Fourier-transformed and brought into a simple tight-binding Majorana model in the momentum space, which looks like
\begin{align}
H_{YL}^{gs, \,per} & = -2 \sum \limits_{\mathbf k \in \frac{1}{2} \text{BZ}}
\psi^{\dagger}( -\mathbf k) 
\begin{bmatrix}
0 & i \, f(\mathbf k) & 0 & 0 & 0 & 0  \\
- i \, f(-\mathbf k) & 0  & 0 & 0 & 0 & 0  \\
0 &  0 & 0 & i \,f (\mathbf k) & 0  & 0 \\
0 &  0 &  - i \, f (-\mathbf k) & 0 & 0  & 0  \\
0 &  0 & 0 & 0 & 0 & i \, f (\mathbf k)  \\
0 &  0 &  0 & 0 & - i \, f (-\mathbf k) & 0  \\
\end{bmatrix}
\psi(\mathbf k) \, ,\nn
\psi(\mathbf k) & = 
\begin{bmatrix}
 a^{(x)} (  \mathbf k) & b^{(x)}(  \mathbf k) & a^{(y)} (  \mathbf k) &
b^{(y)} (  \mathbf k) & a^{(z)} (  \mathbf k) & b^{(z)} (  \mathbf k) 
\end{bmatrix}^T .
\end{align}
Here,
$\lbrace a^{(x)}, \, a^{(y)} , \, a^{(z)} \rbrace $ and
$\lbrace b^{(x)}, \, b^{(y)} , \, b^{(z)} \rbrace $
represent the sets of Majorana fermion operators residing on the triangles located at the A and B sublattices, respectively.
Furthermore,
\begin{align}
f ( \mathbf k) = J_x \, e^{i\, \mathbf k \cdot \mathbf M_1}
+ J_y \, e^{ -i\, \mathbf k \cdot \mathbf M_2} + J_z  \,,
\end{align}
and
\begin{align}
\mathbf M_1 & = \left ( \frac {1} {2}, \, \frac{\sqrt 3} {2} \right )
 \text{ and } \mathbf M_2 =  \left ( \frac {1} {2}, \,- \frac{\sqrt 3} {2} \right )
\end{align}
are the primitive translation vectors of the underlying triangular lattice, with the nearest-neighbor lattice spacing set to unity. The threefold-degenerate energy eigenvalues are given by $ \pm \,2 \, |f(\mathbf k)|$.

Depending on whether $ f ( \mathbf k)$ vanishes at one or multiple points in the $\mathbf k$-space, the
system exhibits a gapless or a gapped phase.
For the Hermitian system, the gapless phase features one Dirac cone in each half of the Brillouin zone for each Majorana species, when the conditions
\begin{align}
\label{eqtriangle}
|J_x| \leq |J_y| + |J_z|\,, \quad
|J_y| \leq |J_y| + |J_z|\,, \text{ and }
|J_z | \leq |J_x | + |J_y | 
\end{align}
are fulfilled. 

\begin{figure}[t]
\centering
\subfigure[]{\includegraphics[width=0.32 \textwidth]{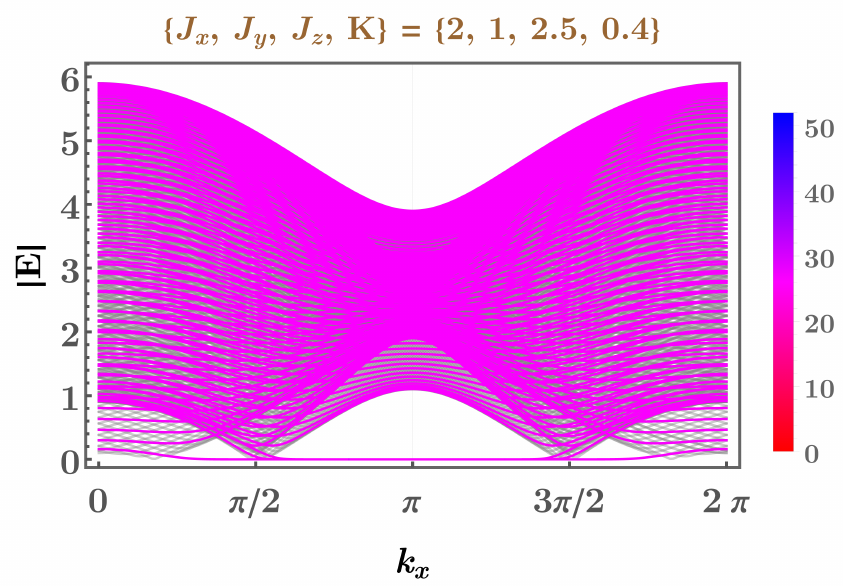}}
\subfigure[]{\includegraphics[width=0.32 \textwidth]{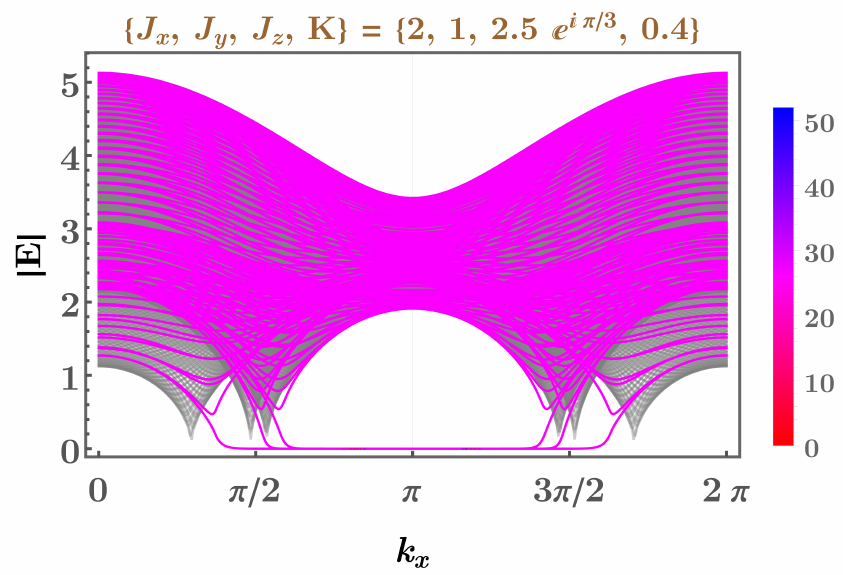}} 
\subfigure[]{\includegraphics[width=0.32 \textwidth]{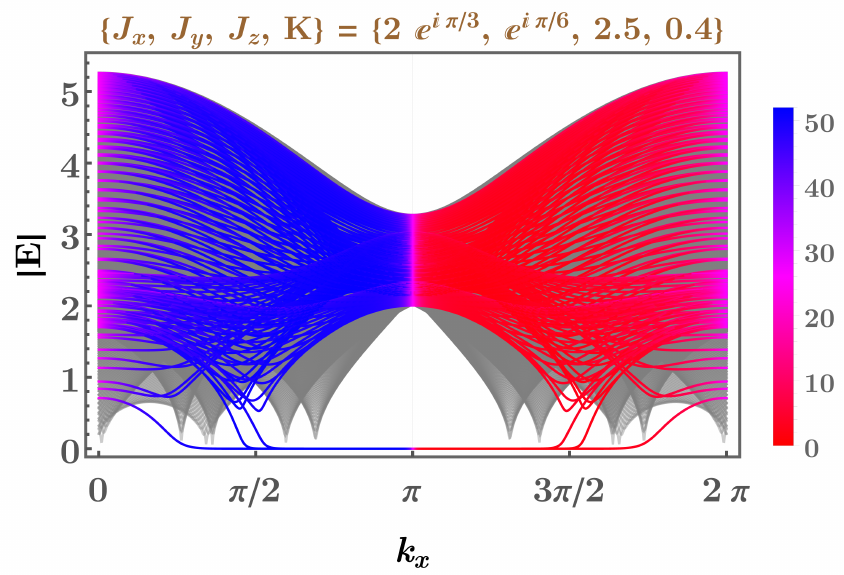}}\\
\subfigure[]{\includegraphics[width=0.32 \textwidth]{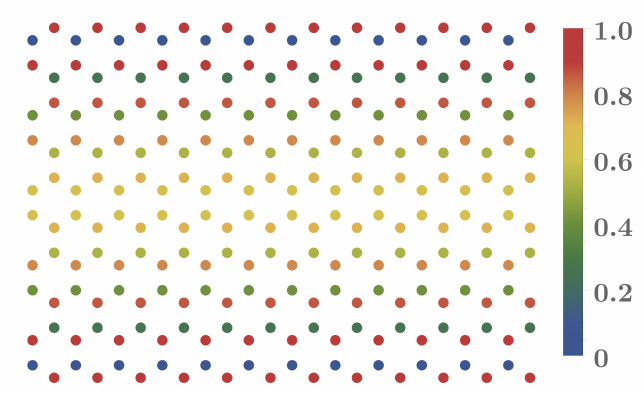}}
\subfigure[]{\includegraphics[width=0.32 \textwidth]{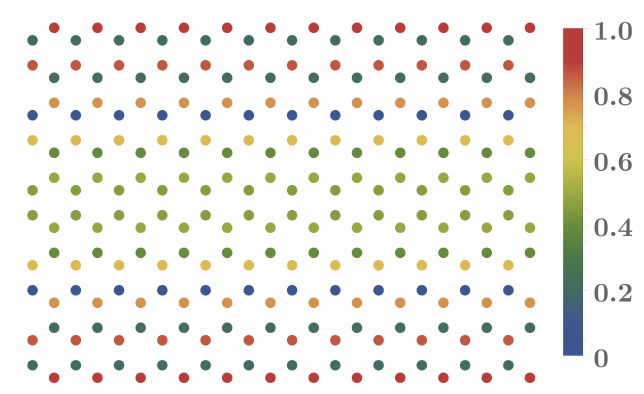}} 
\subfigure[]{\includegraphics[width=0.32 \textwidth]{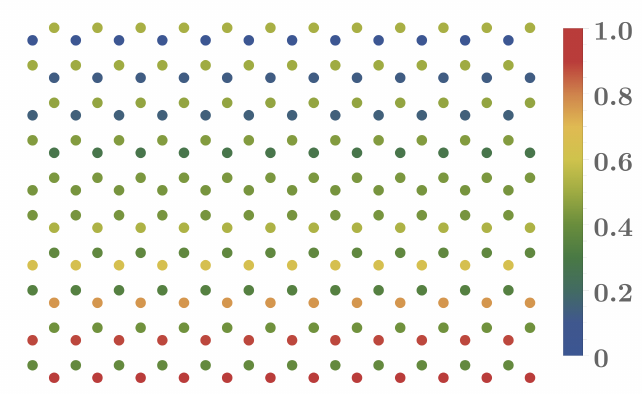}}
\caption{
The absolute values of the eigenvalues (denoted by $|E|$) of the Yao-Lee model plus $K$-interactions [cf. Eq.~\eqref{eqhamKterm}] in the zero-flux sector for OBC (coloured) and PBC (light gray). The OBC spectra are colour-coded,
as shown in the plotlegends, according to the average localization of the eigenstates on the lattice truncated along the $y$-directions. Subfigure (a) represents the Hermitian case with real values of the coupling constants, where the PBC and OBC spectra overlap, except for the three edge modes.
Subfigure (b) has only $J_z$ taking a complex value, with the other two coupling constants remaining real, and does not show NHSE. Subfigure (c) has complex values of $J_x $ and $J_y$, with $J_z$ taken to be real. 
A nonzero phase difference between $J_x$ and $J_y $ gives rise to an NHSE with both the bulk and edge states localizing on one of the two edges. We note that in (a) and (b), since two edge modes are localized at the two opposite boundaries, the color-coding shows the average localization to be the middle of the lattice. All the OBC spectra have been computed using a honeycomb lattice with 52  rows (amounting to a square matrix of dimension $ 6 \times 52 $) and with zigzag edges at the two open boundaries.
Subfigures (d), (e), and (f) illustrate the logarithm of squares of the absolute values of the right eigenvectors ($\psi_R $) for the edge modes on the sublattice points, obtained by setting $ k_x = 2\pi/3$, and normalized such that the minimum and maximum values are zero and one, respectively. These three subfigures correspond to the parameter choices of subfigures (a), (b), and (c), respectively, when the system has zigzag edges.
\label{figK}}
\end{figure}

Nearest-neighbour terms generically arise from the products of the spin ($\boldsymbol \sigma_{j} $) and the orbital ($\boldsymbol \tau_{j}  $) operators \cite{sreejith}. Any nearest-neighbour spin interaction paired with a bond-dependent orbital Ising interaction thus preserves the conservation of the flux operator. Since such a term is quadratic in the itinerant fermions, we can use it break the SO(3) spin-rotational symmetry, thus lifting the eigenvalue degeneracy. In the next two sections, we consider two different types of such nearest-neighbour exchanges, whih are symmetric in the flavour indices.

In the second-last section, we consider a combination of onsite flavour-off-diagonal terms (representing coupling to an external magnetic field) and nearest-neigbour interactions in the form of DMI (which are antisymmetric in the flavour indices).

For all the cases that we investigate in the following sections, we analyze the edge modes for hexagonal lattices which have PBCs along the $x$-direction and OBCs along the $y$-direction (not to be confused with the $x$-type and $y$-type links), with the two open boundaries truncated with zigzag edges. In order to construct the Hamiltonian matrices for the OBC cases, we follow the conventions of Ref.~\cite{manisha}.
In our notations, the $z$-type links are along the $y$-direction and, are, thus perpendicular to the zigzag edges. For comparing the OBC spectra with the PBC spectra, we take the number of layers sandwiched in between the two boundary layers to be even. Since we have three Majorana species, this implies that we must take a square matrix whose dimension is six times an even number, while considering the OBC cases. In what follows, all the eigenvalues are scaled by a factor of 1/2, since a factor of $2$ appears as an overall factor for all the matrices under consideration.

\begin{figure}[t]
\centering
\subfigure[]{\includegraphics[width=0.32 \textwidth]{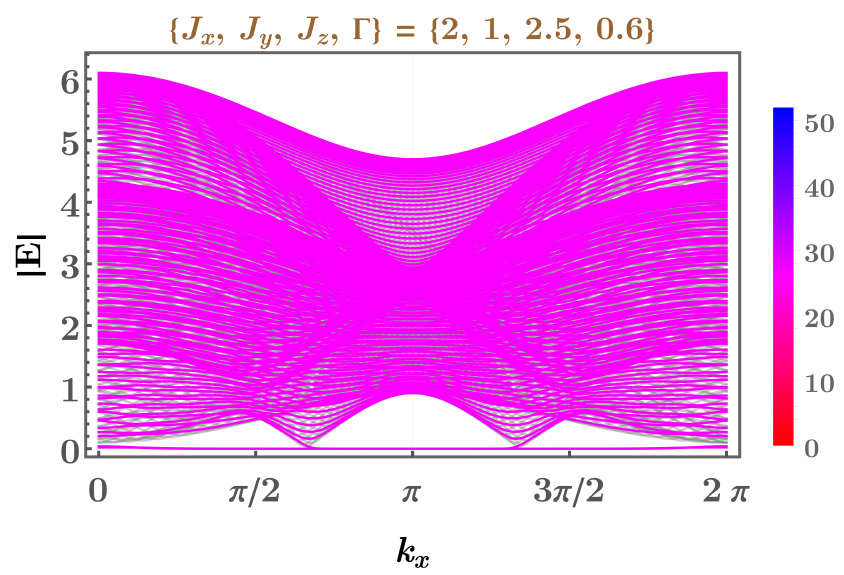}}
\subfigure[]{\includegraphics[width=0.32 \textwidth]{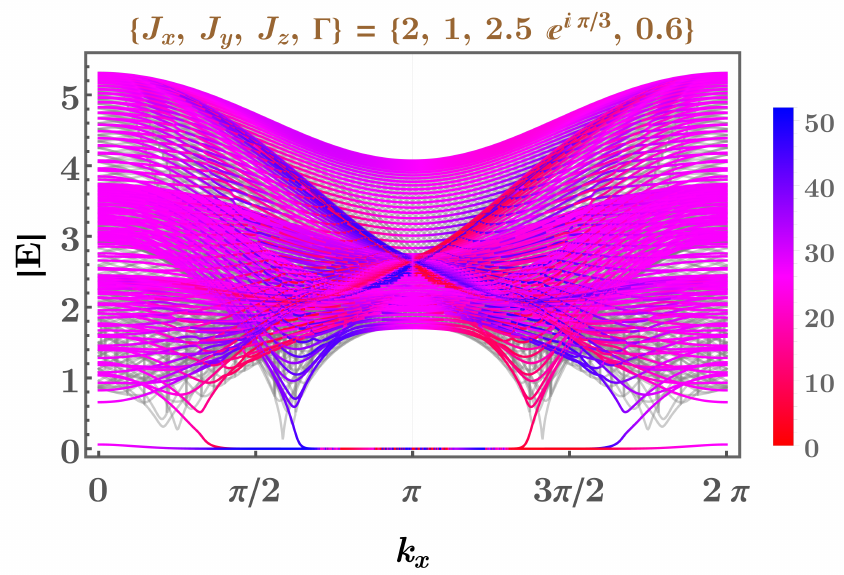}} 
\subfigure[]{\includegraphics[width=0.32 \textwidth]{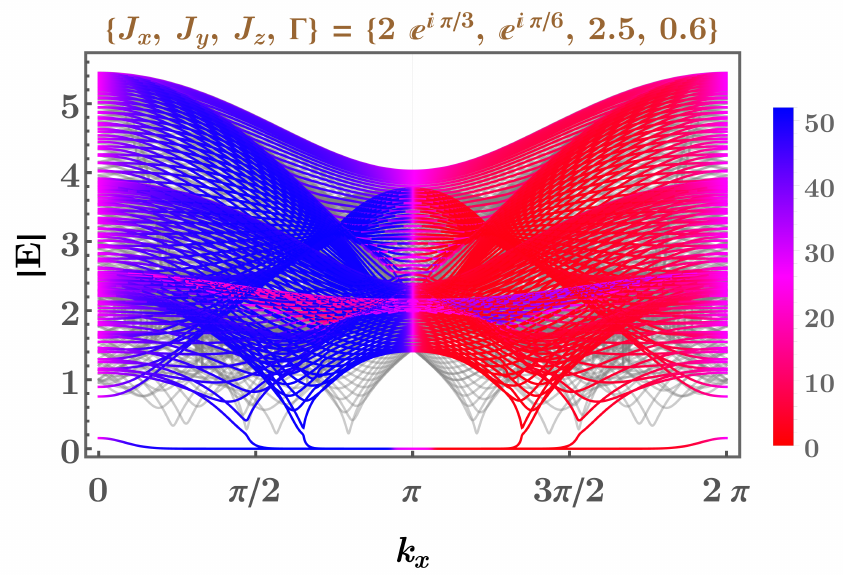}}
\caption{
The absolute values of the eigenvalues (denoted by $|E|$) of the Yao-Lee model plus $\Gamma$-interactions [cf. Eq.~\eqref{eqhamgam}] in the zero-flux sector for OBC (coloured) and PBC (light gray). The OBC spectra are colour-coded,
as shown in the plotlegends, according to the average localization of the eigenstates on the lattice truncated along the $y$-directions. Subfigure (a) represents the Hermitian case with real values of the coupling constants, where the PBC and OBC spectra overlap, except for the three edge modes.
Subfigure (b) has only $J_z$ taking a complex value, with the other two coupling constants remaining real, and shows NHSE. Subfigure (c) has complex values of $J_x $ and $J_y$, with $J_z$ chosen to be real. 
NHSE is observed in both (b) and (c), along with many non-localized eigenstates. We note that in (a) and (b), there exist pairs of edge modes which are localized at the two opposite boundaries and, hence, the color-coding shows the average localization to be the middle of the lattice. All the OBC spectra have been computed using a honeycomb lattice with 52  rows (amounting to a square matrix of dimension $ 6 \times 52 $) and with zigzag edges at the two open boundaries.
\label{figgam}}
\end{figure}

\section{Bond-dependent flavour-diagonal $K$-interaction}
\label{secdiagk}

We add spin-orbital interactions that are bond-dependent in the spin sector, diagonal in the flavour index, and preserve the solvability of the model. In the Majorana representation, they take the form \cite{sreejith}:
\begin{align}
\label{eqhamKterm}
    \mathcal{H}_{K} & =
 i  \, K \sum_{\langle jl \rangle_{\alpha-{\rm links}} }
 u^{\alpha}_{jl} \left [
\delta_{\alpha,  x} \,   c_{j}^{(x)} \, c_{l}^{(x)} 
+  \delta_{\alpha , y} \, c_{j}^{(y)} \,c_{l}^{(y)} 
+  \delta_{\alpha ,  z} \, c_{j}^{(z)} \, c_{l}^{(z)} \right ] ,
\end{align}
with $K$ representing the corresponding coupling strength. 
For a Hermitian system, finite values of $K$ give ground states remaining in the flux-free sector.

In the zero-flux sector, the periodic Hamiltonian, written in the momentum space,
takes the form:
\begin{align}
H_{YL, K}^{gs, \,per} = - 2
\sum \limits_{\mathbf k \in \frac{1}{2} \text{BZ}}
\psi^{\dagger}( -\mathbf k) 
\begin{bmatrix}
0 & i \, A_1(\mathbf k) & 0 & 0 & 0 & 0  \\
- i \, A_1(-\mathbf k) & 0  & 0 & 0 & 0 & 0  \\
0 &  0 & 0 & i \, A_2 (\mathbf k) & 0  & 0 \\
0 &  0 &  - i \, A_2 (-\mathbf k) & 0 & 0  & 0  \\
0 &  0 & 0 & 0 & 0 & i \, A_3 (\mathbf k)  \\
0 &  0 &  0 & 0 & - i \, A_3 (-\mathbf k) & 0  \\
\end{bmatrix}
 \psi(\mathbf k) \, ,
\end{align}
where
\begin{align}
& A_1(\mathbf k) =  f(\mathbf k) +   K\, e^{ i \, \mathbf k \cdot \mathbf M_1} \,, \quad
A_2(\mathbf k) = 
f(\mathbf k) +
  K\, e^{ - i \, \mathbf k \cdot \mathbf M_2}\,, \quad
A_3(\mathbf k) = 
f(\mathbf k) + K \,.
\end{align}
The explicit spectrum is found to be
\begin{align}
& \varepsilon_{K,1} ^ \pm = \pm  \,2\,\left |A_1(\mathbf k) \right  |, \quad 
 \varepsilon_{K,2} ^ \pm = \pm \, 2\, \left | A_2(\mathbf k) \right |, \quad
 \varepsilon_{K,3} ^ \pm = \pm \, 2\, \left | A_3 (\mathbf k) \right | , \nn
\end{align}
To get $ A_1(\mathbf k)$, $ A_2(\mathbf k)$, and $ A_3(\mathbf k)$, we effectively replace $ J_x \rightarrow J_x +K  $, $ J_y \rightarrow J_y +K  $, and $ J_z \rightarrow J_z +K  $, respectively.
Since we have three distinct eigenvalue functions, the question of gaplessness boils down to whether any of the three $A_\eta(\mathbf k)$'s (where $\eta \in \lbrace 1, 2 , 3 \rbrace $) goes to zero. For notational convenience, let us define
\begin{align}
A_\eta(\mathbf k) = J^{(\eta)}_x \, e^{i\, \mathbf k \cdot \mathbf M_1}
+ J^{(\eta)}_y \, e^{ -i\, \mathbf k \cdot \mathbf M_2} + J^{(\eta)}_z\,, 
\end{align}
where
\begin{align}
\mathbf{J} ^{(1)} = \lbrace J_x + K , \, J_y, \, J_z \rbrace \,,\quad
\mathbf{J} ^{(2)} = \lbrace J_x , \, J_y + K , \, J_z \rbrace \,,\quad
\mathbf{J} ^{(3)} = \lbrace J_x , \, J_y, \, J_z+ K \rbrace \,.
\end{align}

\begin{figure}[t]
\centering
{\includegraphics[width= 0.9 \textwidth]{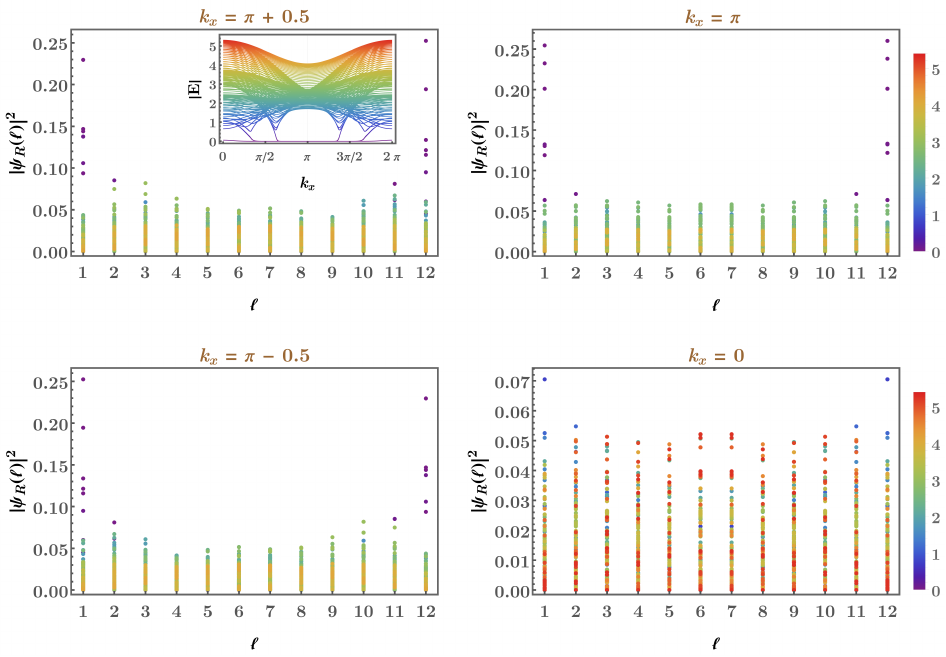}}
\caption{
Plots of the squares of the absolute values of the right eigenvectors ($\psi_R $) for the Yao-Lee model plus $\Gamma$-interactions [cf. Eq.~\eqref{eqhamgam}], with $J_x = 2 $, $J_y = 1 $, $J_z =2.5\, e^{i\,\pi/3}$, and $\Gamma =0. 4 $, versus the lattice sites ($\ell$) of the triangular lattice having OBCs along the $y$-direction. The absolute values of the eigenvalues of the corresponding modes are shown the inset of the first subfigure, reflected in the colour-coding of the plotlegends. The four subfigures depict the localization/delocalization of the eigenmodes for some representative values of $k_x$, as indicated in the plotlabels. The OBCs along the $y$-direction have been implemented by taking a triangular lattice with 10 rows of $z$-links, which translate into 24 sublattice sites and, in total, $3\times 24  $ vertices of the 24 equilateral triangles. 
\label{figgam2}}
\end{figure}

\begin{figure}[t]
\centering
{\includegraphics[width= 0.9 \textwidth]{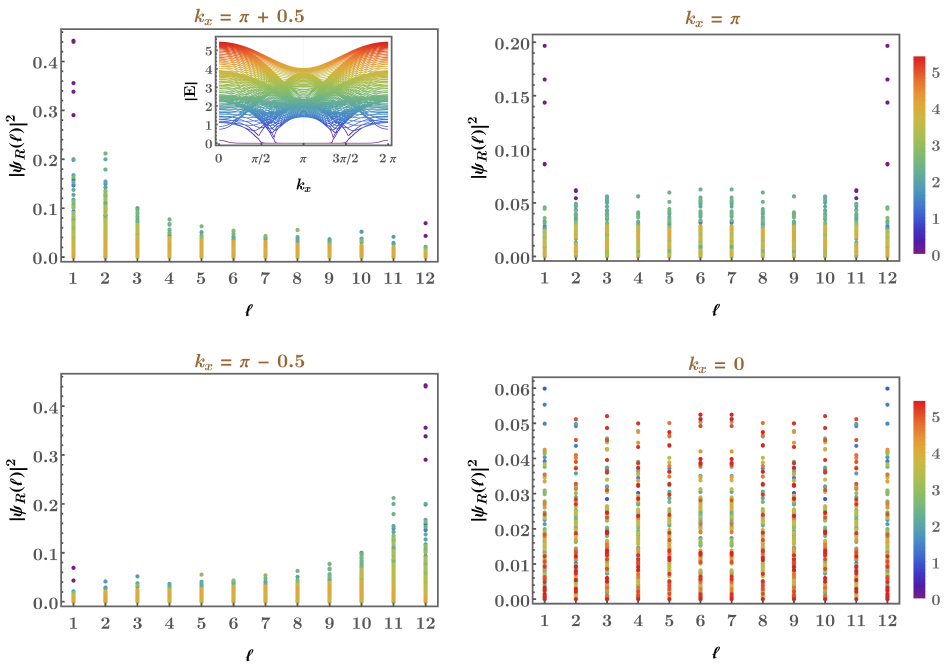}}
\caption{
Plots of the squares of the absolute values of the right eigenvectors ($\psi_R $) for the Yao-Lee model plus $\Gamma$-interactions [cf. Eq.~\eqref{eqhamgam}], with $J_x = 2 \, e^{i\,\pi/3} $, $J_y = e^{i\,\pi/6} $, $J_z =2.5 $, and $\Gamma =0. 4 $, versus the lattice sites ($\ell$) of the triangular lattice having OBCs along the $y$-direction. The absolute values of the eigenvalues of the corresponding modes are shown the inset of the first subfigure, reflected in the colour-coding of the plotlegends. The four subfigures depict the localization/delocalization of the eigenmodes for some representative values of $k_x$, as indicated in the plotlabels. The OBCs along the $y$-direction have been implemented by taking a triangular lattice with 12 rows of $z$-links, which translate into 24 sublattice sites and, in total, $3\times 24 $ vertices of the 24 equilateral triangles.
\label{figgam3}}
\end{figure}

For a system with complex (i.e., non-Hermitian) coupling constants, in general
$A^*_\eta(\mathbf k) \neq A_\eta( -\mathbf k)  $. This gives rise to EPs when either $A_\eta( \mathbf k)$ or $A_\eta( -\mathbf k)$ goes to zero. 
For each value of $\eta$, if the triangle inequality
\begin{align}
|J^{(\eta)}_x| \leq |J^{(\eta)}_y| + |J^{(\eta)}_z|\,, \quad
|J^{(\eta)}_y| \leq |J^{(\eta)}_y| + |J^{(\eta)}_z|\,, \text{ and }
|J^{(\eta)}_z | \leq |J^{(\eta)}_x | + |J^{(\eta)}_y | 
\end{align}
is satisifed, we get two pairs of EPs \cite{kang-emil}.
We parametrize the coupling constants as
$ J^{(\eta)}_x = |J^{(\eta)}_x|\, e^{i\,\phi_{\eta, x}} \, e^{i\,\phi_{\eta, z}} $,
$ J^{(\eta)}_y = |J^{(\eta)}_y|\, e^{i\,\phi_{\eta, y}} \, e^{i\,\phi_{\eta, z}} $, and $ J^{(\eta)}_z = |J^{(\eta)}_z|\, e^{i\,\phi_{\eta, z}} $, such that the overall phase of $  e^{i\,\phi_{\eta, z}} $ is common to all the three $ J^{(\eta)}_\alpha $'s which can be extracted out. We can immediately see that two EPs are located at
\begin{align}
& \lbrace k_{x} , \, k_y \rbrace = \pm
\left \lbrace  
\cos^{-1} 
\left(\frac{| J^{(\eta)}_{y}|^2-|J^{(\eta)}_{x}|^2-|J^{(\eta)}_{z}|^2}
{2\,| J^{(\eta)}_{x}|\, \,   | J_{z}^{(\eta)} |} \right) 
- {\phi}_{\eta, x}, \,
\cos^{-1} 
\left(\frac{| J^{(\eta)}_{x}|^2-|J^{(\eta)}_{y}|^2-|J^{(\eta)}_{z}|^2}
{2\,| J^{(\eta)}_{y}|\, \,| J_{z}^{(\eta)} |} \right) 
- {\phi}_{\eta, y}
\right \rbrace ,\nn
& |J^{(\eta)}_{x}| \sin(  k_x + {\phi}_{\eta, x})  =
 - |J^{(\eta)}_{y}| \sin( k_y + {\phi}_{\eta, y} )\,,
\end{align}
in the reciprocal lattice space, where the second equation fixes the signs in the first. The two EPs are connected by a Fermi arc, and are thus robust against perturbations. 

We find that the right eigenvectors of the system are given by
\begin{align}
\psi^\pm_1 = \lbrace \pm A_1(k), \,  A_1(-k), \,0,\, 0, \, 0, \, 0 \rbrace^T, 
\quad \psi^\pm_2 = \lbrace  0,\,0, \, \pm A_2(k), \,  A_2(-k), \, 0, \, 0 \rbrace^T, 
\quad \psi^\pm_3 = \lbrace  0,\,0 , \, 0, \, 0, \, \pm A_3(k), \,  A_3(-k) \rbrace^T\,.
\end{align}
Hence, even if though there are degeneracy points arising from the eigenvalues for Majoranas of different flavours, they cannot be higher-order EPs. This is because the eigenvectors of the different sectors reside in orthogonal subspaces and can never coalesce, as can be seen from their explicit expressions. In other words, for the points $\mathbf k
= \mathbf  k^{\rm deg} $ where $ \varepsilon_{K,\eta} ^s ( \mathbf  k^{\rm deg}) = \varepsilon_{K,\tilde \eta }^{s'} ( \mathbf  k^{\rm deg}) $ [where of course $\tilde \eta \neq \eta$], we can have two scenarios:
(I) for  $\varepsilon_{K,\eta} ^s ( \mathbf  k^{\rm deg})  \neq \varepsilon_{K,\eta} ^{-s} ( \mathbf  k^{\rm deg}) $, we have normal (i.e., nonsingular) degeneracy points with no EPs;
(II) for  $\varepsilon_{K,\eta} ^+( \mathbf  k^{\rm deg})  = \varepsilon_{K,\eta} ^- ( \mathbf  k^{\rm deg}) $, we have a pair of second-order EPs for each of $\eta$ and $\tilde \eta $.

Basically, the $K$-terms do not bring about any inter-species interaction, although they break the SO(3) symmetry. As a result, each species continues to behave in the way outlines in Ref.~\cite{kang-emil}. But due to differing values of the three $\mathbf J^{(\eta)}$'s, we now have a larger parameter space for the emergence of EPs and NHSE. As derived in Ref.~\cite{kang-emil}, the the general criterion for the appearance of skin effect for the species labelled as $\eta$ is when
\begin{align}
|J_x^{(\eta)} \, e^{i\, k_x a} + J_y^{(\eta)}| \neq 
|J_x^{(\eta)} \, e^{ -i\, k_x a} + J_y^{(\eta)}|\,.
\end{align}
In other words, the skin effect requires a nonzero relative phase between $J_x^{(\eta)}$ and $ J_y^{(\eta)}$, which correspond to the bond directions which are not perpendiculsr to the two parallel zigzag edges.
In Fig.~\ref{figK}, we show the eigenvalue spectra for (a) the Hermitian case with real values of all the coupling constants, (b) a non-Hermitian case with a complex value of $J_z$, and (c) a non-Hermitian case with complex values of $J_x$ and $J_y$ with a relative phase difference. We have chosen a set of parameter values for our plots such that the $\mathbf J$ values are the same as in the figures in Ref.~\cite{kang-emil} --- this helps us to easily compare the modifications we can achieve compared to the single species case of the Kitaev spin liquid.
For all the subfigures, we have taken the same absolute values for all the parameters. We consider the spectra for PBCs on both the $x$- and $y$-directions, which we illustrate in the background in light gray colour. On top of the PBC spectra, we show the spectra for OBCs along the $y$-direction with zigzag edges, while the $x$-direction remains periodic.
We colour-code these OBC spectra according to the average localization of the corresponding eigenstates. We find that the system shows NHSE only when there is a relative phase difference between the three components of $\mathbf J^{(\eta)}$, as argued in Ref.~\cite{kang-emil}.

\section{Bond-dependent flavour-off-diagonal $\Gamma$-interaction}
\label{secgam}

Another set of solvable exchange interactions, which are off-diagonal but symmetric in the flavour indices, is given by \cite{sreejith}
\begin{align}
\label{eqhamgam}
    \mathcal{H}_{\Gamma} & =
 -\, i  \,\Gamma \sum_{\langle jl \rangle_{\alpha-{\rm links}} }
 u^{\alpha}_{jl} \left [
\delta_{\alpha,  x} \left (  c_{j}^{(y)} \, c_{l}^{(z)} 
+  c_{j}^{(z)} \, c_{l}^{(y)} \right) 
+  \delta_{\alpha,  y} \left (  c_{j}^{(z)} \, c_{l}^{(x)} 
+  c_{j}^{(x)} \, c_{l}^{(z)} \right) 
+  \delta_{\alpha,  z} \left (  c_{j}^{(x)} \, c_{l}^{(y)} 
+  c_{j}^{(y)} \, c_{l}^{(x)} \right) 
\right ]
\end{align}
in the Majorana representation, with the coupling constant $\Gamma $. In this case, a closed form expression of the eigenvalues cannot be obtained.

Due to the lack of analytical solutions, we cannot write down closed-form expressions of $\mathbf k$-values where the EPs will arise. We have to find the spectrum numerically. In Fig.~\ref{figgam}, we show the eigenvalue spectra for (a) the Hermitian case with real values of all the coupling constants, (b) a non-Hermitian case with a complex value of $J_z$, and (c) a non-Hermitian case with complex values of $J_x$ and $J_y$ with a relative phase difference. Analogous to Sec.~\ref{secdiagk}, here too we have chosen a set of parameter values for our plots such that the $\mathbf J$ values are the same as in the figures of Ref.~\cite{kang-emil} --- this allows us to easily compare the modifications we can achieve compared to the single species case of the Kitaev spin liquid, as well as the three decoupled species studied in Sec.~\ref{secdiagk}.
For all the subfigures, we have taken the same absolute values for all the parameters. We consider the spectra for periodic boundary conditions (PBCs) on both $x$- and $y$-directions, which we illustrate in the background in light gray colour. On top of the PBC spectra, we show the spectra for OBCs along the $y$-direction with zigzag edges, while the $x$-direction remains periodic.
We colour-code these OBC spectra according to the average localization of the corresponding eigenstates. We find that the system shows NHSE for both cases with complex couplings and, unlike the $K$-interactions studied in the previous section, it does not necessarily require a relative phase difference between $J_x$ and $J_y$. Moreover, in contrast with 
Fig.~\ref{figK}, the non-Hermitian scenarios here harbour a mix of localized and non-localized states at a generic $k_x$. This is seen very clearly in Figs.~\ref{figgam2}
and \ref{figgam3}, which depict the localization of the eigenmodes via the plot of the squares of the absolute values of the right eigenvectors for the complex Hamiltonians, with the parameters used in Fig.~\ref{figgam}(b) and Fig.~\ref{figgam}(c), respectively.

In contrast with the $K$-interaction case, the $\Gamma$-terms induce inter-species interaction, such that the Hamiltonian no longer consists of three dissociated $2\times 2$ blocks. This makes it possible to have EPs which arise from the mixing of the three Majorana species. Consequently, it also follows that the emergence of the NHSE is possible beyond the parameter regimes applicable for Ref.~\cite{kang-emil} and Sec.~\ref{secdiagk}. This can be seen, for example, by comparing Fig.~\ref{figgam}(b) with Fig.~\ref{figK}(b).

\begin{figure}[t]
\centering
\subfigure[]{\includegraphics[width=0.32 \textwidth]{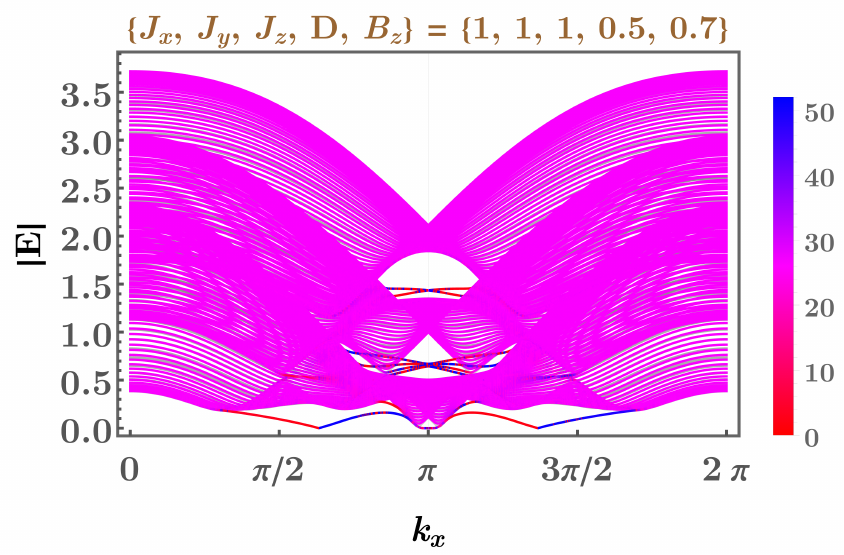}}
\subfigure[]{\includegraphics[width=0.32 \textwidth]{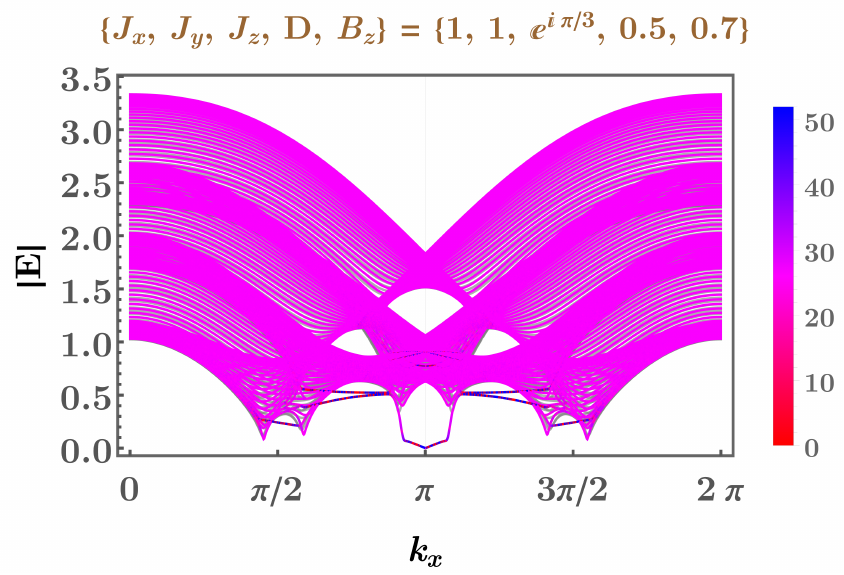}} 
\subfigure[]{\includegraphics[width=0.32 \textwidth]{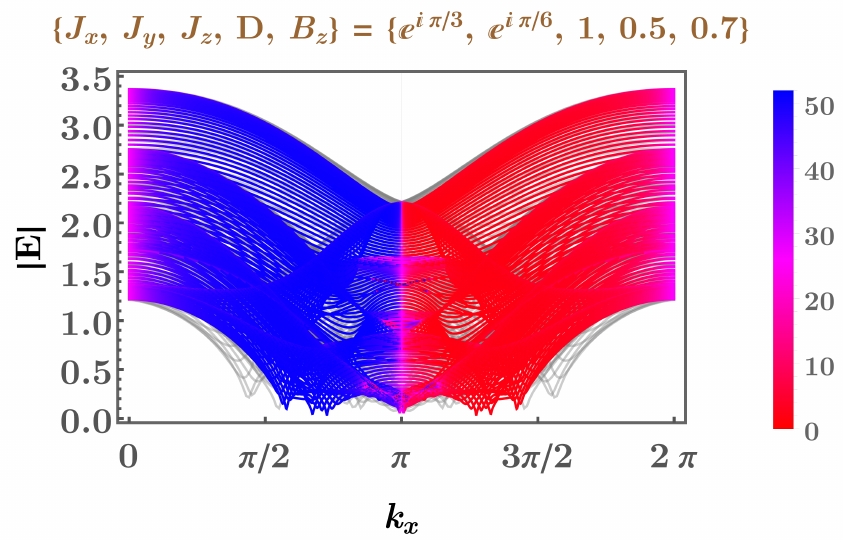}}
\caption{
The absolute values of the eigenvalues (denoted by $|E|$) of the Yao-Lee model coupled with DMI and an external magnetic field [cf. Eq.~\eqref{eqhambdmi}], for OBC (coloured) and PBC (light gray). The OBC spectra are colour-coded, as shown in the plotlegends, according to the average localization of the eigenstates on the lattice truncated along the $y$-directions. Subfigure (a) represents the Hermitian case with real values of the coupling constants, where the PBC and OBC spectra overlap, except for the three edge modes.
Subfigure (b) has only $J_z$ taking a complex value, with the other two coupling constants remaining real, and does not show NHSE for the bulk states. Subfigure (c) has complex values of $J_x$ and $J_y$, with $J_z$ taken to be real.  NHSE for the bulk eigenstates is observed in (c) only, along with a few non-localized eigenstates. All the OBC spectra have been computed using a honeycomb lattice with 52  rows (amounting to a square matrix of dimension $ 6 \times 52 $) and with zigzag edges at the two open boundaries.
\label{figmag}}
\end{figure}

\begin{figure}[t]
\centering
{\includegraphics[width = 0.9 \textwidth]{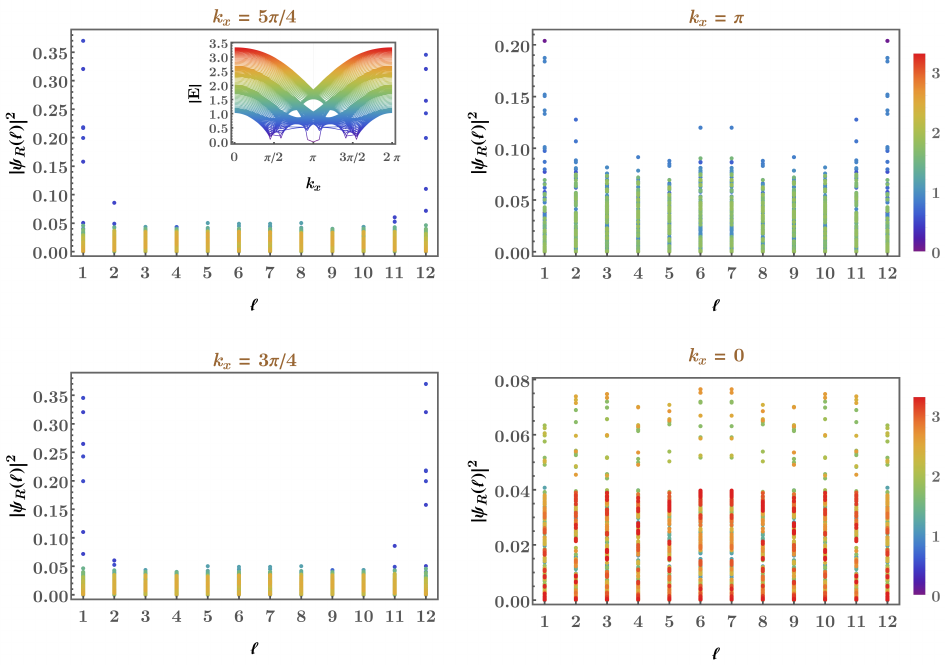}}
\caption{
Plots of the squares of the absolute values of the right eigenvectors ($\psi_R $) for the Yao-Lee model coupled with DMI and an external magnetic field [cf. Eq.~\eqref{eqhambdmi}], setting $J_x =  1 $, $J_y = 1 $, $J_z = e^{i\,\pi/3} $, $D=0.5 $, and $  \pmb{\mathcal B} =0. 7\, \pmb{\hat{z}} $, versus the lattice sites ($\ell$) of the triangular lattice having OBCs along the $y$-direction. The absolute values of the eigenvalues of the corresponding modes are shown the inset of the first subfigure, reflected in the colour-coding of the plotlegends. The four subfigures depict the localization/delocalization of the eigenmodes for some representative values of $k_x$, as indicated in the plotlabels. The OBCs along the $y$-direction have been implemented by taking a triangular lattice with 12 rows of $z$-links, which translate into 24 sublattice sites and, in total, $ 3 \times 24 $ vertices of the 24 equilateral triangles.
\label{figmag2}}
\end{figure}

\begin{figure}[t]
\centering
{\includegraphics[width = 0.9 \textwidth]{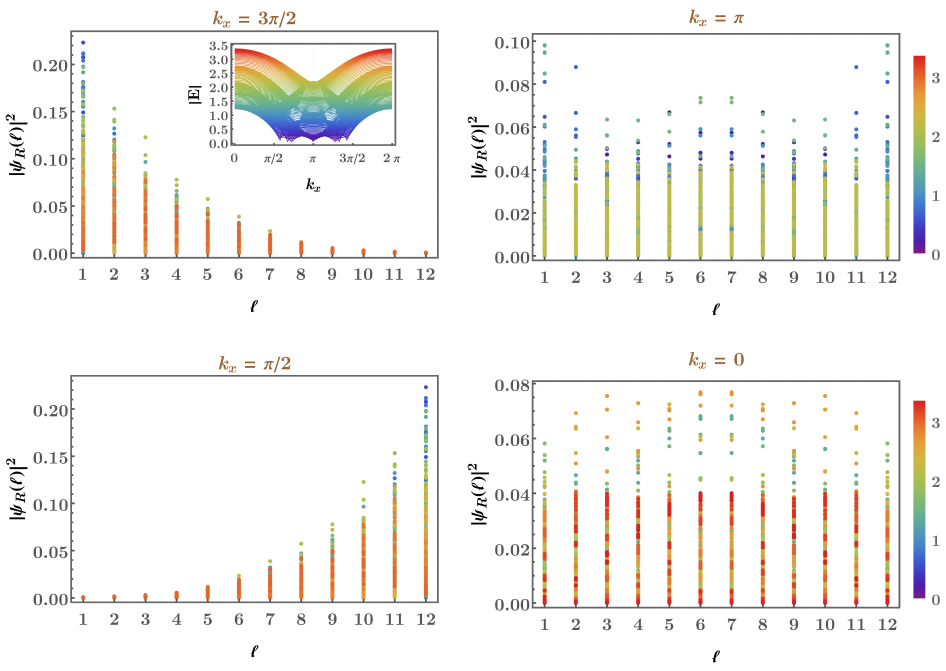}}
\caption{
Plots of the squares of the absolute values of the right eigenvectors ($\psi_R $) for the Yao-Lee model coupled with DMI and an external magnetic field [cf. Eq.~\eqref{eqhambdmi}], setting $J_x =  e^{i\,\pi/3} $, $J_y = e^{i\,\pi/6} $, $J_z = 1 $, $D=0.5 $, and $  \pmb{\mathcal B} =0. 7\, \pmb{\hat{z}} $, versus the lattice sites ($\ell$) of the triangular lattice having OBCs along the $y$-direction. The absolute values of the eigenvalues of the corresponding modes are shown the inset of the first subfigure, reflected in the colour-coding of the plotlegends. The four subfigures depict the localization/delocalization of the eigenmodes for some representative values of $k_x$, as indicated in the plotlabels. The OBCs along the $y$-direction have been implemented by taking a triangular lattice with 12 rows of $z$-links, which translate into 24 sublattice sites and, in total, $3\times 24 $ vertices of the 24 equilateral triangles.
\label{figmag3}}
\end{figure}

\section{DMI in presence of an external magnetic field}
\label{secmag}

In this section, we add SO(3)-symmetry-breaking interaction terms in the form of a spin-orbital DMI as shown below \cite{akram}:
\begin{align}
\label{eqdmi}
    H_{DM}= D \sum \limits_{ \langle jl \rangle_{\alpha-\text{link} }}  
    \left[  \tau_{j}^{(\alpha)} \, \tau_{l}^{(\alpha)} \right ]
\pmb{\hat{\delta}}^{(\alpha, jl)} 
\cdot \left( \pmb{\sigma}_{j}
 \times \pmb{\sigma}_{l} \right), \quad
\pmb{\hat{\delta}}^{(x, jl)} 
= \pmb{\hat{y}} \times \pmb{\hat{z}} \,, \quad
\pmb{\hat{\delta}}^{(y, jl)} 
= -\frac{1}{2} \left( \sqrt 3 \, \pmb{\hat{x}}  + \pmb{\hat{y}} \right)
 \times \pmb{\hat{z}} \,,
\end{align}
in presence of the onsite interactions generated by a uniform external magnetic field $\pmb{\mathcal B} \equiv \lbrace \mathcal B_x, \, \mathcal B_y, \, \mathcal B_z  \rbrace $ through \cite{sreejith,akram}
\begin{align}
\label{eqb}
    H_{B} =  \pmb{\mathcal B}  \cdot \sum_j  \pmb{\sigma}_{j} \,.
\end{align}
Here, $D$ is the DMI coupling constant 
and $\pmb{\hat{\delta}}^{(\alpha, j l)} $ is the DMI vector for the $ \alpha $-link, arising from broken inversion symmetry on the surface \cite{Banerjee_NatPhys2013}. $H_{B}$ represents the sole possible gauge-invariant quadratic onsite terms.

Since $W$ acts only on the orbital degrees of freedom, it commutes with the total Hamiltonian
\begin{align}
H_{\text{mag}} = H_{YL} +  H_{DM} + H_B \,.
\end{align}
Consequently, the eigenstates of $ H_{\text{mag}} $ can still be labelled by the eigenvalues of $W$.

The Hamiltonian $ \mathcal{H}_{\text{mag}} $ in the Majorana representation is related to the original Hamiltonian via $ H_{\text{mag}}=  P \,\mathcal{H}_{\text{mag}}\, P$, where
\begin{align}
\label{eqhambdmi}
\mathcal H_{\text{mag}}  & =  \mathcal{H}_{YL}  + \mathcal{H}_{DM}  + \mathcal{H}_{B} \,,
\nn
    \mathcal{H}_{YL} & =
 i  \sum_{\langle jl \rangle_{\alpha-{\rm links}} }
  J_\alpha \, u^{\alpha}_{jl}
    \left [ c_{j}^{(x)} \, c_{l}^{(x)} +  c_{j}^{(y)} \,c_{l}^{(y)} 
    +   c_{j}^{(z)} \, c_{l}^{(z)} \right ], \nn
\mathcal{H}_{DM} & = i\, D  \sum_{\langle jl \rangle_{\alpha-{\rm links}}
} 
u^{\alpha}_{j l}
\left [ \hat{\delta}^{(\alpha, jl)}_{x}
\left \lbrace c_{j}^{(y)} \,c_{l}^{(z)}- c_{j}^{(z)} \,c_{l}^{(y)} \right \rbrace 
 +  \hat{\delta}^{(\alpha, j l)}_y
 \left \lbrace  c_{j}^{(z)} \,c_{l}^{(x)}- c_{j}^{(x)} \, c_{l}^{(z)}
 \right \rbrace  \right ], \nn
 \mathcal{H}_{B} & =
 i \, \sum_{j} \left [  \mathcal B_x \,c_{j}^{(y)} \, c_{j}^{(z)} 
 +   \mathcal B_y \, c_{j}^{(z)} \, c_{j}^{(x)}
 +  \mathcal B_z \,c_{j}^{(x)}\, c_{j}^{(y)} \right ],
\end{align}
We note that, unlike the $K$- and $\Gamma$-interactions, the DMI terms are both off-diagonal and antisymmetric in the flavour indices.
In Ref.~\cite{akram}, the phase diagram for the case of isotropic Kitaev couplings $ J_x = J_y = J_z = J$ has been chalked out for a Hermitian system.

In the absence of the DMI terms, the effect of $\mathcal{H}_B $ is such that $  |\pmb{\mathcal B} | $ acts as a chemical potential for two of the three sets of the Majorana fermion bands,
while one set of bands remains unaffected by the field in any given flux sector. In particular, for the zero flux sector, the energy bands are captured by the expressions \cite{sreejith}:
\begin{align}
\varepsilon_1^\pm = \pm \, 2\, |f(\mathbf k)| \,, \quad
\varepsilon_2^\pm =  2 \left[ \,
 |\pmb{\mathcal B} | \pm |f(\mathbf k)| \,\right] , \quad
\varepsilon_3^\pm = - 2 \left[ \,  |\pmb{\mathcal B} | \pm |f(\mathbf k)|\, \right ] .
\end{align}
Due to this shift of energy with respect to each other, $\mathcal{H}_B $ causes no nontrivial change in the Dirac cones or the edge-state behaviour of the individual Majorana species. This is the reason why we have added the DMI terms as well. After the addition of the DMI, we can no longer find closed-form analytical expressions for the spectrum.

Using numerical simulations, we show the eigenvalue spectra for (a) the Hermitian case with real values of all the coupling constants, (b) a non-Hermitian case with a complex value of $J_z$, and (c) a non-Hermitian case with complex values of $J_x$ and $J_y$ with a relative phase difference, in Fig.~\ref{figmag}. We have chosen a set of parameter values for our plots such that the original Hermitian version is in the zero-flux sector, according to the phase diagram shown in Ref.~\cite{akram}.
For all the subfigures, we have taken the same absolute values for all the parameters. We consider the spectra for PBCs on both $x$- and $y$-directions, which we illustrate in the background in light gray colour. We overlay them with the colour-coded spectra for OBCs along the $y$-direction with zigzag edges, with the $x$-direction remaining periodic. As before, the colour-coding reflects the average localization of the corresponding eigenstates. 
In Fig.~\ref{figmag}(b), NHSE is not seen for the bulk eigenmodes. Somewhat (but not completely) analogous to the case of $K$-interactions, as well as the case in Ref.~\cite{kang-emil}, we find that most of the bulk eigenmodes show NHSE in Fig.~\ref{figmag}(c), when there is a relative phase difference between the $J_x$ and $J_y$ coupling constants. The difference of this case with the Kitaev model and the $K$-interaction-augmented Yao-Lee model arises from the fact that, for those two cases, we have found that all the bulk modes are localized at the boundaries. In fact, Fig.~\ref{figmag}(c) provides an example where we see extended states with eigenvalues lying within the localized continuum, analogous to the scenarios discussed in Ref.~\cite{maria_emil} in the context of the 1d non-Hermitian Su–Schrieffer–Heeger (SSH) chain.

In order to unambiguously illustrate the localization of the eigenmodes, we show the behaviour of the squares of the absolute values of the right eigenvectors for the complex parameter cases in Figs.~\ref{figmag2} and \ref{figmag3}, with the same parameter values as used in Fig.~\ref{figmag}(b) and Fig.~\ref{figmag}(c), respectively. For the chosen parameter values of Figs.~\ref{figmag}(c) and \ref{figmag3}, we find that the majority of the bulk eigenmodes are localized at one of the boundaries, depending on the value of $k_x$ in question. Albeit, there is a transition of the localization from one boundary to the other as one crosses the values $k_x = 0 $ and $k_x = \pi $.

\section{Summary and outlook}
\label{secsum}

In this paper, we have studied non-Hermitian versions of the Yao-Lee model, supplemented by a variety of SO(3)-symmetry-breaking terms. Due to the presence of three distinct Majorana species, the PBC Hamiltonian itself has a six-band structure, providing a rich platform for the emergence of numerous varieties of non-Hermitian topological phases and exotic NHSEs.
\textcolor{black}{The interplay of the eigenspectra of the three Majorana species gives NHSEs in situations where they would not have emerged if only one Majorana species were present \cite{kang-emil}, or if the three species were decoupled (as are the cases for the pure Yao-Lee model, and the systems with only the $K$-interactions which do not have any inter-species hopping terms).}
We have outlined some of these possibilities by choosing some representative parameter values. 
In particular, we have seen that merely breaking the spectral degeneracy is not enough to give a wider set of parameters which go beyond the conditions found in Ref.~\cite{kang-emil}. Such wider ranges are in fact provided by terms which can lead to a mixing of the three Majorana species. Some practical utilities of the localization phenomena, exhibited by the NHSE, include applications in topological light funneling \cite{weidemann} and stable single-mode lasing \cite{longhi,zhu_lasing}·

In future, it will be worthwhile to formulate an analogue of the biorthogonal polarization \cite{Kunst2018,elisabet,maria_emil, ips-bp}, which has been identified as a real-space invariant whose value is tied to the emergence of the NHSE in 1d non-Hermitian systems, for the 2d and higher dimensional non-Hermitian systems.
\textcolor{black}{
One can count the number of edge modes present in a given parameter regime, and call it a phase transition when one/mode edge modes get/gets delocalized into the bulk, and this can be explicitly found by computing the BP. However, trying to define something analogous to a winding number fails as it does not generically gives the correct quantized value representing the number of edge modes \cite{ips-bp}.
}
Another promising direction involves adding next-nearest-neighbour couplings \cite{yao-lee,sreejith}, and studying the emergent topological phases and NHSE. Last but not the least, it will be interesting to investigate the nature of the edge states when the open boundaries of the honeycomb lattice consist of armchair edges.

\section*{Acknowledgments}

We thank Kang Yang for useful discussions. This research, leading to the results reported in this paper, has received funding from the European Union's Horizon 2020 research and innovation programme, under the Marie Skłodowska-Curie grant agreement number 754340.

\bibliography{bibyl}

\end{document}